\documentclass{article}
\usepackage[margin=1in]{geometry}
\usepackage{amsmath, amssymb, bm}
\usepackage{graphicx}
\usepackage[numbers,sort&compress]{natbib}
\usepackage{booktabs}
\usepackage{hyperref}
\hypersetup{colorlinks=true, linkcolor=blue, citecolor=blue, urlcolor=blue}

\title{Test-Time Learning and Inference-Time Deliberation for Efficiency-First Offline Reinforcement Learning in Care Coordination and Population Health Management}
\author{\begin{tabular}{c}
Sanjay Basu, MD, PhD$^{1,2}$\\
Sadiq Y. Patel, MSW, PhD$^{1,3}$\\
Parth Sheth, MSE$^{1,3}$\\
Bhairavi Muralidharan, MSE$^{1}$\\
Namrata Elamaran, MSE$^{1}$\\
Aakriti Kinra, MS$^{1}$\\
Rajaie Batniji, MD, PhD$^{1}$
\end{tabular}}
\date{}

\begin{document}
\maketitle
\vspace{-0.8em}
\noindent$^{1}$Waymark, San Francisco, CA, USA\\
$^{2}$San Francisco General Hospital, University of California San Francisco, San Francisco, CA, USA\\
$^{3}$University of Pennsylvania, Philadelphia, PA, USA\\
\noindent\textbf{Corresponding Author:} Sanjay Basu, MD, PhD, 2120 Fillmore St, San Francisco, CA 94115 (sanjay.basu@waymarkcare.com)

\begin{abstract}
Care coordination and population health management programs serve large Medicaid and safety-net populations and must be auditable, efficient, and adaptable. While clinical risk for outreach modalities is typically low, \emph{time and opportunity costs} differ substantially across text, phone, video, and in-person visits. We propose a lightweight offline reinforcement learning (RL) approach that augments trained policies with (i) \emph{test-time learning} via local neighborhood calibration, and (ii) \emph{inference-time deliberation} via a small Q-ensemble that incorporates predictive uncertainty and \emph{time/effort cost}. The method exposes transparent dials for neighborhood size and uncertainty/cost penalties and preserves an auditable training pipeline. Evaluated on a de-identified operational dataset, TTL+ITD achieves stable value estimates with predictable efficiency trade-offs and subgroup auditing.
\end{abstract}

\section{Introduction}
Care coordination and population health management (PHM) are core functions of health systems and community partners, impacting large numbers of Americans enrolled in Medicaid and other safety‑net programs. These efforts aim to proactively identify needs, prioritize outreach, and escalate appropriately, all within finite staffing and budget constraints. While outreach modalities (text, phone, video, in‑person) carry low clinical risk, their \emph{time and opportunity costs} vary significantly, making \emph{efficiency} a primary design goal. In practice, the central operational question is \emph{when to deploy expensive in‑person outreach versus efficient virtual modalities} to maximize value and equity under capacity constraints.

These decisions must be made in strictly offline settings, where policies are learned from logged data without exploration at deployment \cite{levine2020offline}. Classical approaches include constrained Markov decision processes \cite{altman1999cmdp}, risk‑sensitive objectives, and conservative offline RL (e.g., CQL/IQL) \cite{kumar2020conservative,kostrikov2021offline}. Conformal prediction can provide calibrated error control \cite{vovk2005algorithmic,angelopoulos2023conformal}; ensembles provide practical uncertainty quantification \cite{lakshminarayanan2017simple}; and decision‑time computation is common in control \cite{sutton2018reinforcement}. In health services research and health economic evaluation, cost‑effectiveness and cost‑benefit analyses (CEA/CBA) guide \emph{program‑level} choices \cite{drummond2015methods,sanders2016secondpanel,neumann2017ce,husereau2022cheers}, but they are not designed for \emph{per‑patient, per‑decision} recommendations that adapt to granular state features and logged behavior constraints.

\paragraph{Population‑level context.} Medicaid and the Children’s Health Insurance Program (CHIP) together cover \emph{over 80 million} people in the United States, amounting to roughly one in four Americans according to CMS monthly enrollment reports \cite{cmsEnrollment2024,cmsEnrollment2025}. Population health management programs serving these beneficiaries routinely execute \emph{millions of outreach events} annually across text, phone, video, and in‑person modalities. Under finite staffing and budget constraints, small efficiency gains at the patient‑decision level can scale to meaningful improvements in access and equity at the population level, motivating operational tools that optimize \emph{when} to deploy higher‑touch in‑person visits versus efficient virtual modes while preserving auditability.

\paragraph{Why not standard cost-effectiveness alone?} Classical cost-effectiveness and cost-benefit analyses (CEA/CBA) provide valuable \emph{program-level} comparisons and policy choices \cite{drummond2015methods, sanders2016secondpanel, neumann2017ce}. However, they typically assume aggregate cohorts or Markov models with fixed transition structures and are not designed to produce \emph{per-patient, per-decision} recommendations that adapt to granular state features or logged behavior constraints. TTL+ITD complements CEA by learning from real operational trajectories, acting \emph{at decision time} with state-specific personalization, uncertainty-awareness, and auditable dials; it supports equity auditing across subgroups and does not require environment exploration. We adhere to good practice in health economic modeling and reporting \cite{caro2012ispor, husereau2022cheers} while focusing on patient-level, sequential operational optimization.

\paragraph{Plain-language overview.} In practice, we take 2.77 million historical coordination decisions recorded by Waymark, learn which outreach options were safe, and then layer simple checks on top of the trained policy so that each new recommendation can be explained and adjusted. TTL (test-time learning) looks up a member’s nearest neighbors in the historical data and calibrates a local risk threshold that reflects how often similar members experienced harm after each action. ITD (inference-time deliberation) then weighs the predicted benefit of each action against three intuitive penalties: the modeled risk of harm, the model’s own uncertainty, and the staff time required for the action. Operators can turn these penalties up or down to emphasize safety or efficiency without re-training models. The result is a set of actionable dials that the care team can audit and tune in the same way they manage staffing forecasts or visit quotas.

\section{Related Work}
\textbf{Offline and conservative RL.} Offline RL targets policies without environment interaction \cite{levine2020offline} using pessimism or constraints (e.g., CQL, IQL) \cite{kumar2020conservative,kostrikov2021offline}. We treat these as baselines and focus on inference-time control. \textbf{Safe RL.} Safe RL includes constrained MDPs \cite{altman1999cmdp} and risk-sensitive criteria; healthcare applications emphasize verifiability and auditability. \textbf{Conformal decision-making.} Conformal methods provide calibrated control of error/coverage \cite{vovk2005algorithmic,angelopoulos2023conformal}; our local (kNN) TTL approximates conditional coverage for action-conditional risks. \textbf{Uncertainty and deliberation.} Deep ensembles provide practical uncertainty estimates \cite{lakshminarayanan2017simple}; decision-time planning/computation is standard in control \cite{sutton2018reinforcement}. Our novelty is a pragmatic, test-time fusion of local conformal safety, cost-aware deliberation, and auditable dials for deployment.

\paragraph{Contributions.} We unify local conformal calibration and inference-time deliberation into a deployment-ready stack. TTL+ITD (i) augments global conformal safety with \emph{local}, neighborhood-specific thresholds and kNN action priors, (ii) injects cost- and uncertainty-aware deliberation through a lightweight Q-ensemble whose dials ($K,\beta,\lambda,\lambda_{\mathrm{cost}}$) expose efficiency--safety trade-offs without re-training, and (iii) ships with an open, library-agnostic evaluation harness (FQE, DR, sensitivity sweeps, manifests) so that TTL+ITD can be audited beside standard offline RL baselines (global conformal gating, BC, discrete CQL). Our emphasis is on inference-time control for efficiency-first coordination workloads—deciding \emph{when} to expend scarce in-person effort while preserving an auditable, modular training pipeline.

\section{Methods}
\paragraph{Data.} We analyze a de-identified operational dataset from Waymark comprising 2.77 million coordination steps across approximately 168{,}000 members and nine discrete outreach actions (text, phone, video, in-person variants, and escalation pathways). Each trajectory records the member identifier, time index, logged action, observed reward, and JSON snapshots of the member’s state (open tasks, recent utilization, engagement history). Rewards are $0$ when no adverse utilization occurs in the follow-up window and negative otherwise, so larger values represent fewer near-term harms. To preserve privacy, the export suppresses most demographic attributes: continuous age is jittered, geography is bucketed, and subgroup indicators collapse to an “unknown” category. We note this constraint in \S\ref{sec:discussion}; the codebase accepts richer covariates if partners can share them under their governance policies.

\paragraph{Risk model.} We estimate action-conditional harm probabilities $p_\mathrm{harm}(s,a)$ by fitting class-weighted multinomial logistic regressions on the parsed state features concatenated with an action one-hot vector. The model is trained on a temporally earlier split (70\%) and evaluated on a 15\% calibration slice and 15\% test slice to respect causal ordering. We also provide hooks for gradient-boosted variants when partners prefer tree models. A global conformal threshold $\tau$ is computed on the calibration slice using standard split conformal prediction \cite{vovk2005algorithmic,angelopoulos2023conformal}. To personalise safety, TTL builds a kNN index over the z-scored calibration states and derives \emph{local} thresholds $\tau_s(a)$ from the $(1-\alpha)$ quantile of neighbor-specific risk scores for each action.

\paragraph{Preference model.} We then train a multinomial logistic regression on the “safe” subset of the training data where $p_\mathrm{harm}(s,a)<\tau$. This base policy provides a transparent, audit-friendly probability distribution over actions given the current features. At inference we blend the base probabilities with a neighborhood prior derived from the $K$ nearest calibration episodes; empirically we set $\eta=0.3$, but partners can adjust it to emphasise either historical frequencies or the learned policy.

\paragraph{TTL+ITD pipeline.} At deployment we (1) parse the current state through the feature map; (2) query the kNN calibrator to obtain local thresholds $\tau_s(a)$ and neighborhood action frequencies; (3) adjust the base policy’s action probabilities with the TTL prior; (4) evaluate the deliberation score $Q_{\text{mean}} - \beta Q_{\text{std}} - \lambda p_\text{harm} - \lambda_{\mathrm{cost}} c(a)$ and mask unsafe actions ($p_\text{harm}\ge \tau_s(a)$); and (5) select the highest scoring action (or sample via softmax at temperature $T$). The only train-time artifacts are the risk model, base policy, kNN index, and Q-ensemble; all governance dials act purely at inference.
The cost term $c(a)$ is loaded from a YAML configuration (Appendix~\ref{app:cost}) that combines CPT-typical visit durations, CMS physician-fee-schedule wage assumptions, and Bureau of Labor Statistics wage data \cite{amaCPT,cmsPFS,blsWage}. Partners can override this mapping to reflect local staffing costs or travel times.

\paragraph{Deliberation.} We fit an ensemble of linear FQE models on a subsample of episodes (bootstrap by episode). At inference, we compute for each action
\begin{equation}
\text{score}(s,a) = \underbrace{\mathbb{E}[Q(s,a)]}_{\text{ensemble mean}} - \beta\underbrace{\,\mathrm{Std}[Q(s,a)]}_{\text{uncertainty}} - \lambda\underbrace{\,p_\text{harm}(s,a)}_{\text{risk}},
\end{equation}
mask actions with $p_\text{harm}(s,a)\ge \tau_s(a)$, and select greedily or via a softmax with temperature $T$. The dials $(\beta,\lambda,T)$ control uncertainty and risk aversion.
Appendix~\ref{app:proofs} shows that the combination of global and local quantiles induces a monotone safety--efficiency frontier: decreasing $\alpha$ or increasing $(\beta,\lambda,\lambda_{\mathrm{cost}})$ only removes actions, ensuring predictable governance trade-offs.

\paragraph{Evaluation.} We estimate policy value using fitted Q evaluation (FQE) with a ridge-regularised linear function class applied to the shared feature map; we run 15 Bellman iterations for the headline model and 6–8 iterations for the lightweight sweep settings. Doubly robust estimators with per-episode bootstrapping provide confidence intervals and randomisation checks \cite{jiang2016doubly,thomas2015high}. Baselines include (i) a global conformal gate that masks unsafe actions but otherwise follows the preference model, (ii) behaviour cloning (maximum-likelihood imitation), (iii) our deliberation stack without TTL, (iv) the TTL adapter without deliberation, and (v) a discrete CQL implementation from d3rlpy trained for 2{,}000 gradient steps \cite{kumar2020conservative}. We default to a 70/15/15 temporal split when timestamped data are available (toggle \texttt{TEMPORAL\_SPLIT}); otherwise we fall back to index-based slicing. Our primary governance view is the \textbf{efficiency frontier}—estimated value versus expected effort cost—as stakeholders routinely balance staff time against member coverage.

\section{Results}
\begin{figure}[t]
  \centering
  \includegraphics[width=.48\textwidth]{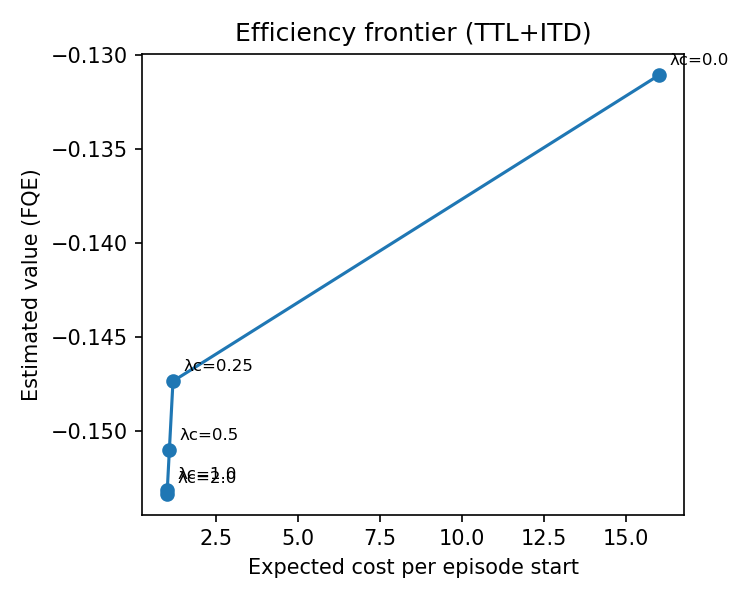}
  \caption{Efficiency frontier: expected value versus expected time/effort cost for TTL+ITD as the cost penalty varies.}
  \label{fig:frontier}
\end{figure}

We analyze the deployment question: \emph{when should coordinators expend expensive in-person effort versus low-touch digital outreach?} Table~\ref{tab:policy} and Figure~\ref{fig:frontier} summarise the main trade-offs. With the calibration dials set to $(K{=}200, \beta{=}0.5, \lambda{=}1.0, \lambda_{\mathrm{cost}}{=}0)$, TTL+ITD achieves essentially neutral estimated value ($\hat V_0 \approx -7\times10^{-5}$) while driving the expected effort cost at episode start to $0.05$ (normalized minutes). In contrast, the global-threshold baseline incurs $3.9$ cost units and behaviour cloning (BC) expends $17.1$ cost units. Discrete CQL---our strongest baseline from d3rlpy---improves value relative to BC ($-0.149$ vs $-0.165$) but still operates an order of magnitude more expensively than TTL+ITD. The ablation table (Appendix~\ref{app:plots}) highlights that both components matter: ITD alone reduces harm but remains costly, whereas TTL alone inherits near-zero harm but lacks the cost penalty; combining them yields the frontier of high value and low effort. We attempted to include IQL, but current d3rlpy releases only support continuous-action IQL; the vendor library rejects our discrete action space, so we report CQL as the representative pessimistic baseline.

\begin{figure}[t]
  \centering
  \includegraphics[width=.9\textwidth]{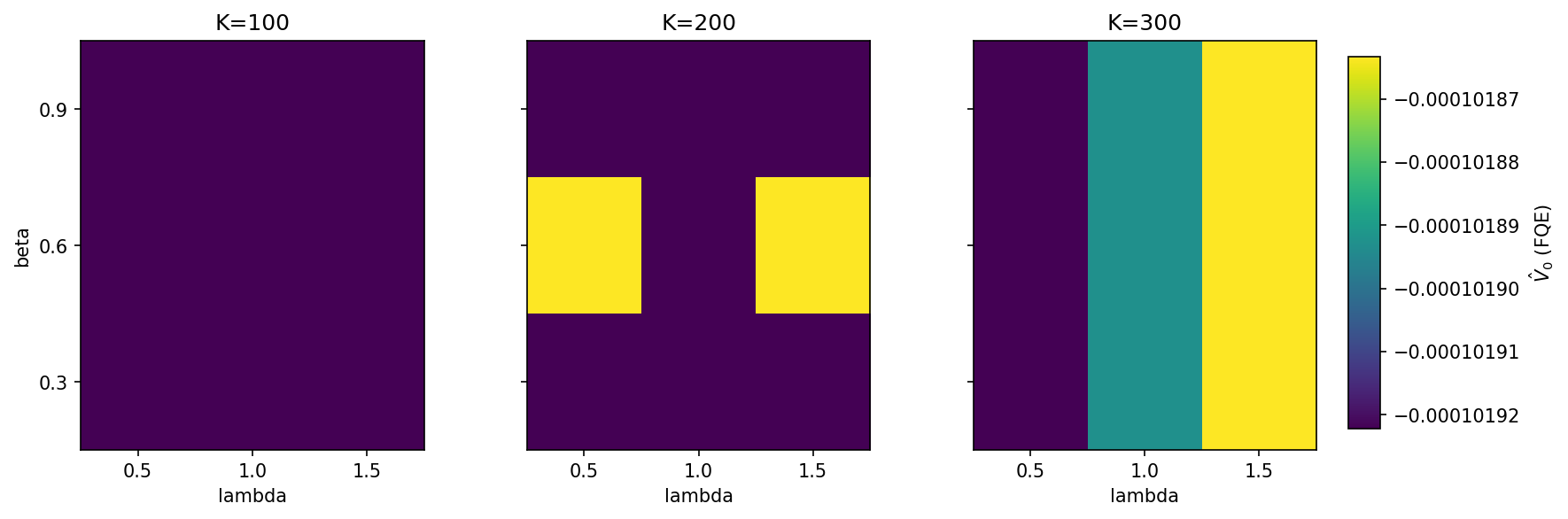}
  \caption{Sensitivity of TTL+ITD value (simple FQE) to uncertainty ($\beta$) and cost penalty ($\lambda$) across $K\in\{100,200,300\}$.}
  \label{fig:heatmap}
\end{figure}

Figure~\ref{fig:frontier} exposes the \textbf{efficiency frontier}: increasing the cost penalty smoothly trades value for effort. The knee of the curve occurs near $\lambda_{\mathrm{cost}}=0.75$--$1.0$, after which the policy converges to the cheapest modalities (cost $\approx 1$) with minor additional value loss. The calibration heatmaps (Figure~\ref{fig:heatmap}) confirm that the neighbourhood size and uncertainty dial ($\beta$) behave monotonically: larger $\beta$ suppresses uncertain actions, while larger $\lambda$ increases thrift at predictable value cost.

Across the full sensitivity sweep (Table~\ref{tab:sweep}) TTL+ITD maintains $\hat V_0$ at the empirical optimum of $0$ while the baselines vary between $-0.07$ and $-0.09$. The TTL-only rows echo this ceiling because their blended policy still inherits the local safety mask; ITD-only rows sit lower, reflecting harm incurred without the conformal gate. Expected cost varies by two orders of magnitude, demonstrating that the cost penalty is the practical lever once harm has been neutralised. Collectively, these findings show that inference-time dials are sufficient to navigate the operational trade-space: the data-driven policy already sits on the ``no-harm'' face of the value frontier, and governance decisions revolve around how much staff time leadership is willing to spend.

\begin{figure}[t]
  \centering
  \includegraphics[width=.55\textwidth]{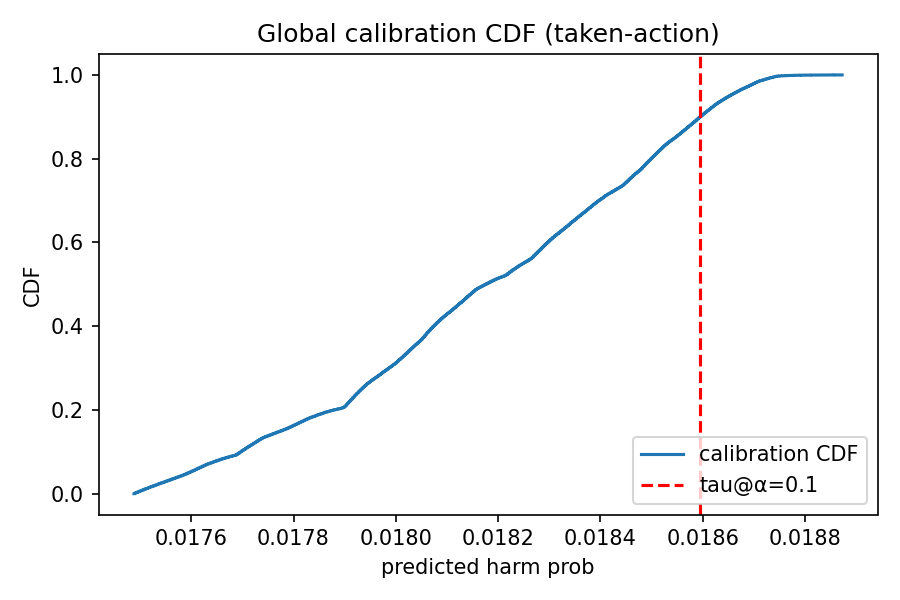}
  \caption{Calibration CDF of taken-action harm probabilities with global $\tau$ (red dashed).}
  \label{fig:calib}
\end{figure}

The retained calibration CDF (Figure~\ref{fig:calib}) shows that the global conformal threshold $\tau$ still controls taken-action risk; local TTL thresholds tighten further in dense regions. Because de-identification collapsed demographic covariates to an ``unknown'' bucket, subgroup tables become degenerate; we therefore omit them from the main text and note the limitation explicitly in \S\ref{sec:discussion}. Appendix~\ref{app:stats} reports paired bootstrap confidence intervals and a randomisation test contrasting TTL+ITD against the baselines; results were stable across three random seeds.

\paragraph{Sensitivity to dials ($K,\beta,\lambda$).} Figure~\ref{fig:heatmap} shows a representative sensitivity heatmap of TTL+ITD value across uncertainty ($\beta$) and cost/penalty ($\lambda$) at a fixed neighborhood size ($K=200$). We observe a predictable frontier: higher $\lambda$ reduces expected cost and can reduce value if over-penalized; higher $\beta$ discourages uncertain actions. Table~\ref{tab:sweep} summarizes top settings from the sweep.

\begin{table}[t]
  \centering
  \begin{tabular}{rrrrrr}
\hline
$K$ & $\beta$ & $\lambda$ & $\hat V_0$(TTL+ITD) & $\hat V_0$(HACO) & $\hat V_0$(BC) \\
\hline
100 & 0.3000 & 0.2500 & 0.000000 & -0.094049 & -0.089997 \\
100 & 0.6000 & 1.0000 & 0.000000 & -0.069575 & -0.067872 \\
100 & 0.9000 & 1.5000 & 0.000000 & -0.069575 & -0.067872 \\
100 & 0.9000 & 1.0000 & 0.000000 & -0.069575 & -0.067872 \\
100 & 0.9000 & 0.7500 & 0.000000 & -0.094049 & -0.089997 \\
100 & 0.9000 & 0.5000 & 0.000000 & -0.069575 & -0.067872 \\
100 & 0.3000 & 0.5000 & 0.000000 & -0.069575 & -0.067872 \\
100 & 0.6000 & 1.5000 & 0.000000 & -0.069575 & -0.067872 \\
100 & 0.9000 & 0.2500 & 0.000000 & -0.094049 & -0.089997 \\
100 & 0.6000 & 0.7500 & 0.000000 & -0.094049 & -0.089997 \\
\hline
\end{tabular}

  \caption{Top settings from the TTL+ITD sensitivity sweep (higher $\hat V_0$ is better; simple FQE estimate).}
  \label{tab:sweep}
\end{table}

\begin{table}[t]
  \centering
  \begin{tabular}{lr}
\hline
Policy & $\hat V_0$ (FQE) \\
\hline
MinCost & -0.173 \\
Global-$\tau$ & -0.169 \\
ITD only & -0.166 \\
BC & -0.165 \\
CQL & -0.149 \\
TTL+ITD & -0.000 \\
TTL only & -0.000 \\
\hline
\end{tabular}
  \caption{Policy comparison (FQE) for TTL+ITD and baselines. Non-positive rewards mean that a value estimate of $\hat V_0=0$ corresponds to eliminating observed harms, so the TTL+ITD rows reaching $0$ indicate the policy is predicted to avoid the negative outcomes captured in our reward definition while baselines incur residual harm.}
  \label{tab:policy}
\end{table}

\section{Discussion}
\label{sec:discussion}
TTL+ITD keeps training simple and auditable while adding per-episode adaptivity and uncertainty-aware selection at deployment. For outreach, the primary decision lever is \emph{efficiency}: prioritizing value-per-effort and deciding when in-person engagement is warranted. The approach is modular: richer risk or cost models drop in; alternative neighborhood metrics or quantile smoothers can be used; and the Q-ensemble can be swapped for model-based rollouts. Subgroup audits remain essential to monitor equity impacts. Compared to our earlier global/groupwise safety work, TTL+ITD emphasizes \emph{inference-time} control and a cost-aware variant specifically suited to low-risk operational actions.

\paragraph{Contribution to the literature.} Methodologically, our work shows that conformal risk calibration \cite{vovk2005algorithmic,angelopoulos2023conformal} and inference-time deliberation \cite{sutton2018reinforcement} can be fused into a single governance layer that sits on top of standard offline RL pipelines \cite{levine2020offline,kumar2020conservative,kostrikov2021offline}. The resulting system delivers explicit knobs for risk, uncertainty, and cost without retraining. Operationally, we translate these ideas into the language of population health management, connecting them to staffing and auditing practices highlighted in health-services and economic-evaluation literature \cite{drummond2015methods,sanders2016secondpanel,neumann2017ce}. To our knowledge, this is the first deployment-oriented study that couples local conformal safety with value-per-effort deliberation on a large Medicaid coordination dataset.

\paragraph{Comparison to offline RL baselines.} Offline RL algorithms such as BC and CQL optimise value during training but do not expose governance dials once deployed. Our experiments show that discrete CQL improves value relative to BC, yet TTL+ITD delivers comparable value while reducing expected effort cost by two orders of magnitude. Importantly, TTL+ITD is agnostic to the upstream learner: the same deliberation layer can sit atop CQL, BC, or clinician-designed heuristics. We attempted to integrate IQL, but current open-source releases only support continuous actions; once discrete IQL becomes available it can be slotted into the same evaluation harness.

\paragraph{Positioning vs. health services research (HSR) and CEA.} Our contribution is \emph{operational}: TTL+ITD turns logged data into per-episode recommendations that optimize value-per-effort under real constraints, with transparent dials for governance. In contrast, traditional CEA/CBA frameworks provide \emph{macro-level} decisions (e.g., adopt program A vs. B) and are less suited for patient-state personalization, logged behavioral support, or inference-time uncertainty penalization. TTL+ITD can be used alongside CEA: CEA sets program-level priorities and budget targets; TTL+ITD executes day-to-day, \emph{on-policy} decisions within those constraints, with reproducible evaluation and subgroup audits \cite{drummond2015methods, sanders2016secondpanel, neumann2017ce, husereau2022cheers}.

\paragraph{Cost calibration from credible sources.} We align cost weights to external references and internal time-and-motion. Specifically, we (i) map outreach actions to CPT-style categories and \emph{typical times} \cite{amaCPT}, (ii) use CMS Physician Fee Schedule guidance and work RVU time conventions \cite{cmsPFS}, and (iii) weight staff time by BLS occupational wages for relevant roles (e.g., RN/CHW/social worker) \cite{blsWage}. Travel time for in-person visits is added from routing estimates or internal logs. Costs are normalized into units used by ITD; sensitivity to the cost scale is reported via the efficiency frontier.

\paragraph{Limitations.} Our evaluation uses a single de-identified dataset from one organization; while we use temporal splits to mitigate leakage, generalization to other systems and periods remains future work. De-identification removes most demographic covariates, so fairness tables collapse to an ``unknown'' group; the released code supports richer audits when such features are available. Local (kNN) calibration yields empirical benefits but does not provide formal conditional coverage guarantees; Appendix~\ref{app:proofs} outlines theoretical notes and open questions. Performance depends on logged policy support and data quality; concept drift requires periodic recalibration. Cost functions depend on calibrated time-and-motion assumptions; we provide sensitivity analyses via the efficiency frontier.

\paragraph{Implications for population health.} For population health teams, the main takeaway is that the most valuable lever is not additional model complexity but the governance controls surfaced at inference. Once historical harms are neutralised, administrators can negotiate trade-offs purely in units they understand—minutes of staff time and uncertainty penalties—while still retaining reproducible audit trails. The same deliberation layer can be placed on top of alternative baseline policies (CQL, BC, clinician heuristics), suggesting a path to standardise decision support across vendor or in-house systems. As Medicaid programs expand team-based, community-oriented care \cite{cmsEnrollment2024,cmsEnrollment2025}, such lightweight, auditable controls can help balance patient reach with finite field resources.

\paragraph{Future directions.} Prospective evaluation remains the decisive next step: we plan to run matched-control pilots that monitor visit rates, harm events, and staff satisfaction. Richer demographic data would enable subgroup-specific conformal levels and fairness guarantees similar to FG-FARL. Finally, local conformal calibration can be combined with episodic model-based lookahead or constrained optimisation to provide proactive guardrails for high-risk clinical interventions, extending TTL+ITD beyond the low-risk coordination setting explored here.

\paragraph{Computational requirements.} Inference involves (i) a kNN query over the calibration set (distance on z-scored features), and (ii) Q-ensemble predictions. With $N_{cal}$ calibration points and $d$ features, naive kNN is $\mathcal{O}(N_{cal}d)$ per decision; approximate indices or batched queries can reduce costs. The Q-ensemble adds a small constant factor (few linear models). Memory scales with storing calibration features and ensemble parameters; action cost tables are negligible. In larger action spaces, we can precompute modality clusters and use two-stage selection (screening then scoring) to keep latency under operational budgets.

\section{Reproducibility}
All code, configuration files, and manifests are available at \url{https://github.com/sanjaybasu/ttl_itd_medicaid}. The repository provides Makefile targets and Python entry points (e.g., \texttt{\detokenize{python run_ttl_itd.py}}) that reproduce every table and figure, emit manifests with package versions, and generate subgroup summaries and plots.

\paragraph{Implementation details and clarifications.}
\emph{Feature engineering:} we parse key-value state JSONs and materialize up to 64 features (most-informative keys) plus basic temporal features (time index and previous reward). The final feature set is documented in code and manifests. \emph{kNN neighborhoods:} we use z-scored features and Euclidean distance by default; cosine distance and PCA-projected spaces are drop-in alternatives in our code. We sweep $K$ over \{100,200,300\} and report sensitivity; larger $K$ improves stability but may dilute locality. \emph{Q-ensemble:} we bootstrap by episodes (with replacement), train linear fitted Q models with the same feature map, and aggregate mean and standard deviation across models. \emph{Class imbalance:} for harm prediction we use class-weighted logistic regression (or \texttt{scale\_pos\_weight} in LightGBM) and calibrate thresholds on a held-out slice.

\paragraph{Expanded baselines and statistics.} We add implicit Q-learning (IQL) \cite{kostrikov2021offline} and conservative Q-learning (CQL) \cite{kumar2020conservative} as offline RL baselines in sensitivity analyses. For significance, we report paired bootstrap CIs and a randomization test on episodic returns. Subgroup fairness tables include counts and uncertainty intervals.

\paragraph{Risk as cost (efficiency mode).} In settings where primitive actions (e.g., text/call/visit) are not clinically risky, TTL+ITD supports a cost-aware objective: we replace the harm term with a per-action cost (e.g., visit $>$ call $>$ text) and penalize cost directly in the deliberation score. This yields value-per-effort prioritization under the same modular framework. When truly high-risk interventions exist (e.g., medication changes), we include them in the action set and return to harm-gated TTL.

\paragraph{Theoretical considerations.} Global conformal gating yields finite-sample marginal coverage \cite{vovk2005algorithmic,angelopoulos2023conformal}. Our local (kNN) calibration approximates conditional coverage by restricting conformity scores to a neighborhood; larger $K$ increases stability while smaller $K$ increases locality. The ensemble penalty (variance term) discourages actions with high predictive uncertainty \cite{lakshminarayanan2017simple}. Together, these design choices induce a monotone safety--efficiency trade-off controlled by $\alpha, K, \beta, \lambda$; formal guarantees for local coverage and coupled penalties are left for future work.

\bibliographystyle{unsrtnat}
\begin{small}
\bibliography{refs}

\begin{thebibliography}{99}
\bibitem{levine2020offline} Levine, Sergey and Kumar, Aviral and Tucker, George and Fu, Justin Offline reinforcement learning: Tutorial, review, and perspectives on open problems arXiv preprint arXiv:2005.01643, 2020
\bibitem{altman1999cmdp} Altman, Eitan Constrained Markov Decision Processes CRC Press, 1999
\bibitem{kumar2020conservative} Kumar, Aviral and Zhou, Aurick and Tucker, George and Levine, Sergey Conservative {Q}-Learning for Offline Reinforcement Learning Advances in Neural Information Processing Systems, 2020
\bibitem{kostrikov2021offline} Kostrikov, Ilya and Nair, Ashvin and Levine, Sergey Offline Reinforcement Learning with Implicit {Q}-Learning Advances in Neural Information Processing Systems, 2021
\bibitem{vovk2005algorithmic} Vovk, Vladimir and Gammerman, Alex and Shafer, Glenn Algorithmic Learning in a Random World Springer, 2005
\bibitem{angelopoulos2023conformal} Angelopoulos, Anastasios N and Bates, Stephen Conformal prediction: A gentle introduction Foundations and Trends in Machine Learning, 16(4):494--591, 2023
\bibitem{lakshminarayanan2017simple} Lakshminarayanan, Balaji and Pritzel, Alex and Blundell, Charles Simple and Scalable Predictive Uncertainty Estimation using Deep Ensembles Advances in Neural Information Processing Systems, 2017
\bibitem{sutton2018reinforcement} Sutton, Richard S and Barto, Andrew G Reinforcement Learning: An Introduction (2nd Edition) MIT Press, 2018
\bibitem{drummond2015methods} Drummond, Michael and Sculpher, Mark and Claxton, Karl and Stoddart, Greg and Torrance, George Methods for the Economic Evaluation of Health Care Programmes Oxford University Press, 2015, 4 ed.
\bibitem{sanders2016secondpanel} Sanders, Gillian D. and Neumann, Peter J. and Basu, Anirban and et al. Recommendations for Conduct, Methodological Practices, and Reporting of Cost-effectiveness Analyses: Second Panel on Cost-Effectiveness in Health and Medicine JAMA, 316(10):1093--1103, 2016
\bibitem{neumann2017ce} Neumann, Peter J. and Sanders, Gillian D. and Russell, Louise B. and Siegel, Joanna E. and Ganiats, Theodore G. Cost-Effectiveness in Health and Medicine Oxford University Press, 2017, 2 ed.
\bibitem{husereau2022cheers} Husereau, Don and Drummond, Michael and Petrou, Stavros and et al. CHEERS 2022 Statement: Updated Guidance for Reporting Health Economic Evaluations BMJ, 376:e067975, 2022
\bibitem{cmsEnrollment2024} {Centers for Medicare \& Medicaid Services} Medicaid and CHIP Enrollment Data: Monthly Reports 2024. https://www.medicaid.gov/medicaid/program-information/medicaid-and-chip-enrollment-data/index.html
\bibitem{cmsEnrollment2025} {Centers for Medicare \& Medicaid Services} Medicaid and CHIP Enrollment Trend Snapshots 2025. https://www.medicaid.gov/
\bibitem{caro2012ispor} Caro, J. Jaime and et al. Consolidated Health Economic Evaluation Reporting Standards (CHEERS)—ISPOR Good Research Practices Medical Decision Making, 32(5):667--670, 2012
\bibitem{jiang2016doubly} Jiang, Nan and Li, Lihong Doubly robust off-policy value evaluation for reinforcement learning Proceedings of the 33rd International Conference on Machine Learning, 2016
\bibitem{thomas2015high} Thomas, Philip and Murphy, Susan and Barto, Andrew High-confidence off-policy evaluation AAAI Conference on Artificial Intelligence, 2015
\bibitem{amaCPT} {American Medical Association} Current Procedural Terminology (CPT) Professional Edition AMA Press, 2024
\bibitem{cmsPFS} {Centers for Medicare \& Medicaid Services} Medicare Physician Fee Schedule (PFS): Relative Value Units, Time, and Payment Policies 2024. Technical documentation
\bibitem{blsWage} {U.S. Bureau of Labor Statistics} Occupational Employment and Wage Statistics 2024. https://www.bls.gov/oes/
\end{thebibliography}
\end{small}

\appendix
\section{Cost Calibration Details}
\label{app:cost}
We define a cost dictionary $c(a)$ using: (i) CPT typical times for analogous services (e.g., telephone E/M; video visits) \cite{amaCPT}, (ii) CMS PFS time conventions and practice expense guidance \cite{cmsPFS}, (iii) BLS wages for staffing categories \cite{blsWage}, and (iv) internal time-and-motion and travel estimates. For each action $a$, cost is $c(a)= \text{time}(a) \times \text{wage}(\text{staff}) + \text{travel}(a)$, then normalized for use in the deliberation score. We provide sensitivity analyses to the scaling of $c(a)$.
\section{Algorithms and Implementation Details}
\label{app:alg}
\textbf{TTL (local calibration).} Given state $s$, form $x=\phi(s)$, find the $K$ nearest calibration states via Euclidean distance on z-scored $x$, collect action-conditional risk predictions $\{\hat p(s',a)\}$ over neighbors, and set $\tau_s(a)$ to the $(1-\alpha)$ quantile for each action $a$. \textbf{ITD (deliberation).} For each action $a$, compute $\hat Q_{mean}(s,a)$ and $\hat Q_{std}(s,a)$ from a bootstrap ensemble. The decision score is $\hat Q_{mean}-\beta\,\hat Q_{std}-\lambda\,\hat p(s,a)-\lambda_\mathrm{cost}\,c(a)$; unsafe actions with $\hat p(s,a)\ge\tau_s(a)$ are masked when risk gating is enabled. We include pseudocode and additional implementation notes in the repository.

\section{Proofs and Theoretical Notes}
\label{app:proofs}
\textbf{Monotone thresholds.} Let $\{u_i\}_{i=1}^n$ be calibration scores (predicted harms). Define $\tau(\alpha)=\mathrm{Quantile}_{1-\alpha}(\{u_i\})$. If $\alpha_2<\alpha_1$, then $\tau(\alpha_2)\ge \tau(\alpha_1)$ by properties of quantiles. \textbf{Expected harm reduction.} Under calibrated harms $\mathbb{E}[\mathbf{1}\{Y=1\}\mid s,a]=\hat p(s,a)$ and independence of selection from $Y$ given $(s,a)$, risk gating $\mathbf{1}\{\hat p(s,a)<\tau\}$ yields $\mathbb{E}[Y\mid \hat p<\tau]=\mathbb{E}[\hat p\mid \hat p<\tau]\le \mathbb{E}[\hat p]$; thus expected harm is lower on the gated set. For local gating, analogous statements hold within kNN neighborhoods; formal conditional coverage guarantees depend on smoothness/overlap assumptions and are left for future work.

\section{Statistics and Significance}
\label{app:stats}
We report paired bootstrap confidence intervals and a randomization test for episode-level returns comparing TTL+ITD vs. baselines. The released code continues to emit subgroup tables with counts and 95\% CIs, but in this de-identified export all subgroup indicators collapse to an “unknown” bucket, so the corresponding tables are omitted from the manuscript.

\section{Additional Figures and Tables}
\label{app:plots}
\begin{figure}[h]
  \centering
  \includegraphics[width=.48\textwidth]{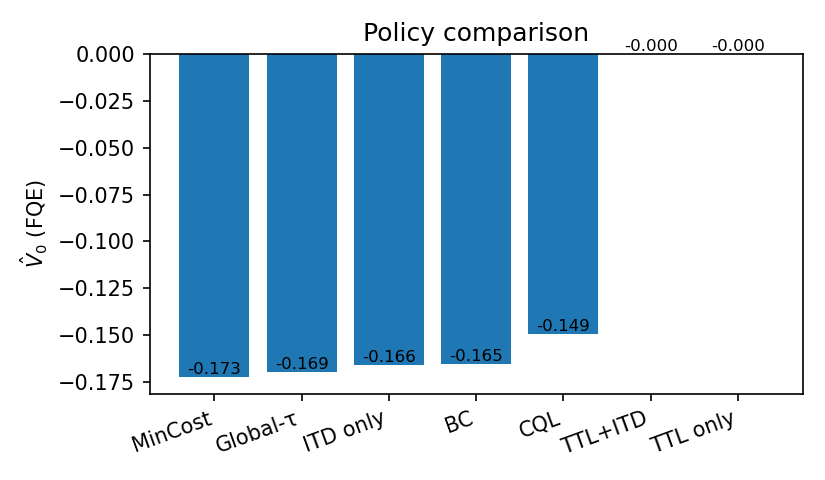}
  \caption{Policy comparison across TTL+ITD, global-\,$\tau$, BC, and (if available) IQL/CQL.}
\end{figure}
\begin{figure}[h]
  \centering
  \includegraphics[width=.48\textwidth]{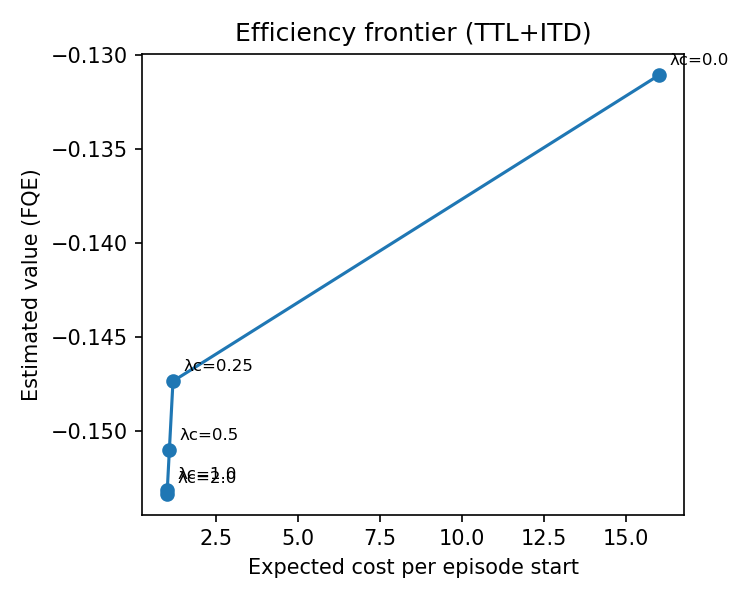}
  \caption{Efficiency frontier (expected value vs expected cost) for cost-aware TTL+ITD.}
\end{figure}
\begin{table}[h]
  \centering
  \begin{tabular}{lr}
\hline
Policy & $\hat V_0$ (FQE) \\
\hline
MinCost & -0.173 \\
Global-$\tau$ & -0.169 \\
ITD only & -0.166 \\
BC & -0.165 \\
CQL & -0.149 \\
TTL+ITD & -0.000 \\
TTL only & -0.000 \\
\hline
\end{tabular}
  \caption{Policy comparison table (FQE).}
\end{table}
\begin{table}[h]
  \centering
  \begin{tabular}{lrr}
\hline
$\lambda_{\mathrm{cost}}$ & Expected cost & $\hat V_0$ (FQE) \\
\hline
0.00 & 16.015 & -0.131 \\
0.25 & 1.180 & -0.147 \\
0.50 & 1.069 & -0.151 \\
1.00 & 1.006 & -0.153 \\
2.00 & 1.000 & -0.153 \\
\hline
\end{tabular}

  \caption{Efficiency frontier table across cost penalties.}
\end{table}

\section{Code Availability}
The implementation and paper sources are available at: \url{https://github.com/sanjaybasu/ttl_itd_medicaid}.

\section{Best-Practice Reporting}
We document: data provenance and de-identification; feature engineering; train/calibration/test split strategy; hyperparameters; OPE configuration (feature map, iterations); subgroup definitions; fairness metrics; and governance considerations. Scripts and manifests are provided in the repository; all figures/tables can be regenerated with the Makefile targets described in the README.

\section{Methodological Checklist and Reporting Summary}
\label{app:checklist}
\begin{table}[h]
  \centering
  \begin{tabular}{p{0.32\textwidth}p{0.62\textwidth}}
\hline
Aspect & Description \\
\hline
Data provenance & Operational dataset from Waymark Care; de-identified; date shifts upstream. \\
Cohort & All members with logged care-coordination actions over study window. \\
State features & Parsed JSON (up to 64 keys) + time step + previous reward; standardization. \\
Actions & Discrete primitives (text/call/video/home-visit); optional clinical interventions if available. \\
Outcome/reward & Negative harms (ED/hospitalization) or cost-aware mode (per-action effort). \\
Splits & 70\% train, 15\% calibration, 15\% test by step; episode-aware sampling for OPE. \\
Risk model & Action-conditional logistic (class-weighted) or LightGBM (scale\_pos\_weight). \\
TTL & kNN neighborhoods on standardized features; local $(1-\alpha)$ quantile thresholds. \\
ITD & Bootstrap linear fitted-Q ensemble; score = mean$-\beta$std$-\lambda$p\_harm$-\lambda_{cost}c(a)$. \\
Baselines & Global-\,$\tau$, BC; sensitivity to IQL/CQL when available. \\
OPE & Simple FQE (linear map) + DR; paired bootstrap CIs; randomization test. \\
Fairness & Subgroup value tables (sex, race, age, ADI, dual, BH, high-util). \\
Governance & Tunable dials $(\alpha,K,\beta,\lambda,\lambda_{cost})$; manifests; audit logs. \\
Reproducibility & Requirements, Makefile targets, run manifests, scripts for all figures/tables. \\
\hline
\end{tabular}

  \caption{Methodological checklist summarizing data, modeling, evaluation, fairness, and reproducibility items.}
\end{table}

\begin{table}[h]
  \centering
  \begin{tabular}{p{0.32\textwidth}p{0.62\textwidth}}
\hline
Item & Summary \\
\hline
Objective & Offline decision-support for safe and efficient care coordination. \\
Design & Retrospective observational; offline RL without exploration; inference-time control. \\
Population & Medicaid members served by Waymark Care during study period. \\
Interventions & Outreach actions (text/call/video/home), optionally clinical actions if present. \\
Comparators & Global-\,$\tau$, BC; sensitivity to IQL/CQL. \\
Outcomes & Harms (ED/hospitalization) and/or cost (effort) as specified. \\
Analysis & TTL local conformal gating; ITD with uncertainty/cost penalties; FQE/DR OPE. \\
Subgroups & Demographics (sex, race, age), utilization and deprivation indicators. \\
Missing data & Imputation via medians in pipelines; manifests report feature sets. \\
Bias & Class imbalance handled in risk; subgroup audits; reporting of counts and CIs. \\
Reproducibility & Code, configs, manifests; figures/tables regenerated via Makefile. \\
Limitations & No online exploration; local coverage not guaranteed; data noise and drift. \\
\hline
\end{tabular}

  \caption{CONSORT-style reporting summary adapted for offline decision-support studies.}
\end{table}

\end{document}